\documentclass[10pt,pra,aps,notitlepage,superscriptaddress,reprint]{revtex4-2}

\usepackage[hyperindex,breaklinks,colorlinks=true]{hyperref}

\usepackage{bm}

\usepackage[utf8]{inputenc}
\usepackage[T1]{fontenc}
\usepackage[english]{babel}  

\usepackage{tikz}

\usepackage{colortbl}
\usepackage{tabu,diagbox}

\usepackage{pgfplots}
\usepackage{lipsum}
\usepackage{youngtab}
\usepackage{subfigure}

\usepackage{times}
\usepackage{dsfont}

\usepackage{amssymb,amsmath,amsfonts,amsthm}
\usepackage{mathtools}
\usepackage{verbatim}
\usepackage{enumerate}
\usepackage{graphicx}

\usepackage[capitalise]{cleveref}
\usepackage{bbm}
\usepackage{booktabs}

\newcommand{\appsection}[1]{\let\oldthesection\thesection
  \renewcommand{\thesection}{Appendix \oldthesection}
  \section{#1}\let\thesection\oldthesection}

\def\Dbar{\leavevmode\lower.6ex\hbox to 0pt
{\hskip-.23ex\accent"16\hss}D}

\newcommand{\nc}{\newcommand}

\nc{\cA}{{\cal A}} \nc{\cB}{{\cal B}} \nc{\cC}{{\cal C}}
\nc{\cD}{{\cal D}} \nc{\cE}{{\cal E}} \nc{\cF}{{\cal F}}
\nc{\cG}{{\cal G}} \nc{\cH}{{\cal H}} \nc{\cI}{{\cal I}}
\nc{\cJ}{{\cal J}} \nc{\cK}{{\cal K}} \nc{\cL}{{\cal L}}
\nc{\cM}{{\cal M}} \nc{\cN}{{\cal N}} \nc{\cO}{{\cal O}}
\nc{\cP}{{\cal P}} \nc{\cQ}{{\cal Q}} \nc{\cR}{{\cal R}}
\nc{\cS}{{\cal S}} \nc{\cT}{{\cal T}} \nc{\cU}{{\cal U}}
\nc{\cV}{{\cal V}} \nc{\cW}{{\cal W}} \nc{\cX}{{\cal X}}
\nc{\cZ}{{\cal Z}}

\def\i{\iota}

\def\affa{\affiliation{Shenzhen SpinQ Technology Co., Ltd., Shenzhen, China}}
\def\affb{\affiliation{College of Physics and Electronic Engineering \& Center for Computational Sciences,  Sichuan Normal University, Chengdu, China}}
\def\affc{\affiliation{Department of Physics, The Hong Kong University of Science and Technology, Clear Water Bay, Kowloon, Hong Kong}}
\def\affd{\affiliation{Department of Physics, Tsinghua University, Beijing, China}}
\def\affe{\affiliation{Department of Physics,
Southern University of Science and Technology, Shenzhen, China}}
\def\afff{\affiliation{Institute for Quantum Computing, University of Waterloo, Waterloo, Ontario, Canada}}

\begin{document}

\title{SpinQ Gemini: a desktop quantum computer for education and research}

\author{Shi-Yao Hou}
\affa
\affb
\affc

\author{Guanru Feng}
\affa

\author{Zipeng Wu}
\affc
\affd

\author{ Hongyang Zou}
\affa

\author{Wei Shi}
\affa

\author{Jinfeng Zeng}
\affc

\author{Chenfeng Cao}
\affc 

\author{Sheng Yu}
\affa 

\author{Zikai Sheng}
\affa 

\author{Xin Rao}
\affa

\author{Bing Ren}
\affa 

\author{Dawei Lu}
\affe 

\author{Junting Zou}
\affa 

\author{Guoxing Miao}
\email{guo-xing.miao@uwaterloo.ca}
\afff
\affa

\author{Jingen Xiang}
\email{jxiang@spinq.cn}
\affa

\author{Bei Zeng} 
\email{zengb@ust.hk}
\affc

\begin{abstract}
SpinQ Gemini is a commercial desktop quantum computer designed and manufactured by SpinQ Technology. 
It is an integrated hardware-software system. The first generation product with two qubits
was launched in January 2020. The hardware is based 
on NMR spectrometer, with permanent magnets providing $\sim 1$ T magnetic field.
SpinQ Gemini operates under room temperature ($0$-$30^{\circ}$C),
highlighting its lightweight (55 kg with a volume of $70\times 40 \times 80$ cm$^3$), cost-effective (under $50$k USD),
and maintenance-free. SpinQ Gemini aims to provide real-device experience for quantum computing education for K-12 and at the college
level. It also features quantum control design capabilities that benefit the researchers studying quantum control and quantum noise.
Since its first launch, SpinQ Gemini has been shipped to institutions in Canada, Taiwan and Mainland China. This paper introduces 
the system of design of SpinQ Gemini, from hardware to software. We also demonstrate
examples for performing quantum computing tasks on SpinQ Gemini, including one task for 
a variational quantum eigensolver of a two-qubit Heisenberg model.
The next generations
of SpinQ quantum computing devices will adopt models of more qubits, advanced control functions for researchers with comparable cost,  as well as simplified 
models for much lower cost  (under $5$k USD) for K-12 education. 
We believe that low-cost portable quantum computer products will facilitate hands-on experience for teaching quantum computing
at all levels, well-prepare younger generations of students and researchers for the future of quantum technologies.
\end{abstract}

\date{\today}

\pacs{03.65.Ud, 03.67.Dd, 03.67.Mn}

\maketitle

\section{Introduction}

\begin{figure*}[!htbp]
\includegraphics[scale=0.7]{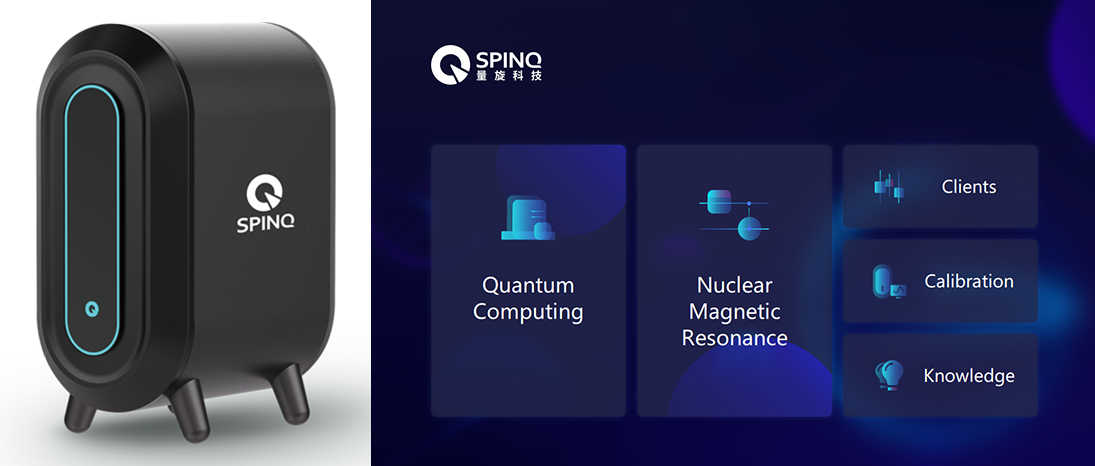}
\caption{The photo of Gemini (left) and the homepage of \textsc{SpinQuasar} (right). The Gemini connects with a personal computer (PC) installed with \textsc{SpinQuasar}. \textsc{SpinQuasar} provides an interface for the users to manipulate the desktop quantum computer.}
\label{photo}
\end{figure*}

SpinQ Gemini is a commercial desktop quantum computer designed and manufactured by SpinQ Technology~\cite{patent1},
and the first generation product with two qubits was launched in January 2020.
It is an integrated hardware-software system as shown in~Fig. \ref{photo}: the left figure shows the exterior look
of the device, with a dimension of $70\times 40 \times 80$ cm$^3$, and a weight of $55$kg; the right figure shows the user interface
software SpinQuasar.

The hardware part of Gemini is based on NMR spectrometer. 
NMR was among the very 
first systems developed for quantum computing~\cite{Corypps,Gershenfeld97bulkspin-resonance,ChuangBulk,CORY199882,Rayonequbit,Cory2000,nmrrev2005}.
Despite its limitations on scalability,
a lot of pioneer research and techniques for quantum computing were first demonstrated in NMR systems~\cite{nmr1,nmr2,nmr3,nmr4,nmr5,nmr6,Fourier,ErrorCorrection,ErrorCorrection5,ShorExp,LINDEN199861,geometricNMR,adiabaticNMR}. 
Notably, many quantum control techniques developed in NMR can be readily applied to other quantum computing platforms~\cite{CP,CPMG,nmrrev2005,grape1,grape2,strongly,composite,compositeChuang,compositeJones,RobustDDSuter} . 

Traditional NMR quantum computing is performed on commercial spectrometers with a superconducting magnet. Those spectrometers are expensive (almost 1 million USD), large (can be as high as $\sim$3 meters), and need to work in specially designed labs. They also require regular liquid nitrogen and helium refills for maintenance. These issues on cost, weight, volume and extreme physical conditions also in general exist on other systems for quantum computing,
making them hard to be accessible for users in real life, but instead with only possible access on cloud, such as IBM Q~\cite{cross2018ibm}, IonQ~\cite{monroeionq} and Rigetti~\footnote{https://www.rigetti.com/}.
As an example, superconducting qubits need to work in dilution fridges which usually cost almost 1 million USD.  Similar as the NMR systems, they require special lab conditions and are not portable. Furthermore, special training is needed for operations of dilution fridges~\cite{kjaergaard2020superconducting,krantz2019quantum,wendin2017quantum,gu2017microwave,you2006superconducting,
you2011atomic,wendin2007quantum}. 

With the development of permanent magnet technology in recent years~\cite{magnet1}, it is possible to bring down the size and cost of NMR spectrometers~\cite{magnet2,spinsolve,oxford,nanalysis,anasazi}.
This then makes the NMR technology an ideal choice for building portable quantum computers. 
By using a permanent magnet providing 1 T magnetic fields, 
SpinQ Gemini highlights its lightweight (55 kg with a volume of 70$\times$40$\times$80 cm$^3$) and cost-effective (under $50$k USD) features, and maintenance-free, making it portable almost like a desktop PC.

Customised quantum algorithm circuit design and programming are supported on SpinQ Gemini using its software SpinQuasar (Fig.~\ref{photo}).
SpinQ Gemini also provides demonstrations of $>10$ famous quantum algorithms, such as Deutsch algorithm~\cite{DJalgorithm}, Grover algorithm~\cite{grover1996fast,Groverlong}, and HHL algorithm~\cite{harrow2009quantum}. It also has build-in teaching examples for quantum mechanic, such as Rabi oscillation observation and decoherence time measurement. 
Gemini not only provides a very friendly platform for non-specialists who aim to learn quantum computing basics and quantum programming fast, but also serves as a powerful tool for quantum computing related research.

In this paper, we introduce the system of the first generation SpinQ Gemini. In Sec. II, we discuss the system design, from hardware to software. In Sec. III, we discuss how to perform quantum computing with SpinQ Gemini. We give two concrete examples of quantum computing tasks performed on SpinQ Gemini, one on the measurement of geometric phase of mixed state in Sec. IV, and the other on a variational quantum eigensolver for a two-qubit Heisenberg model in Sec. V. A discussion on future plans of next generations products will follow in Sec. VI.

We believe that low-cost portable quantum computer products will facilitate hands-on experience for teaching quantum computing
at all levels, well-prepare younger generations for the future of quantum technologies. It will also be accessible to a wider range of researchers to operate under real world conditions for quantum computers, benefiting them for further studies on quantum control and quantum noise.

\section{System}
\begin{figure*}[!htb]
\includegraphics[scale=0.12]{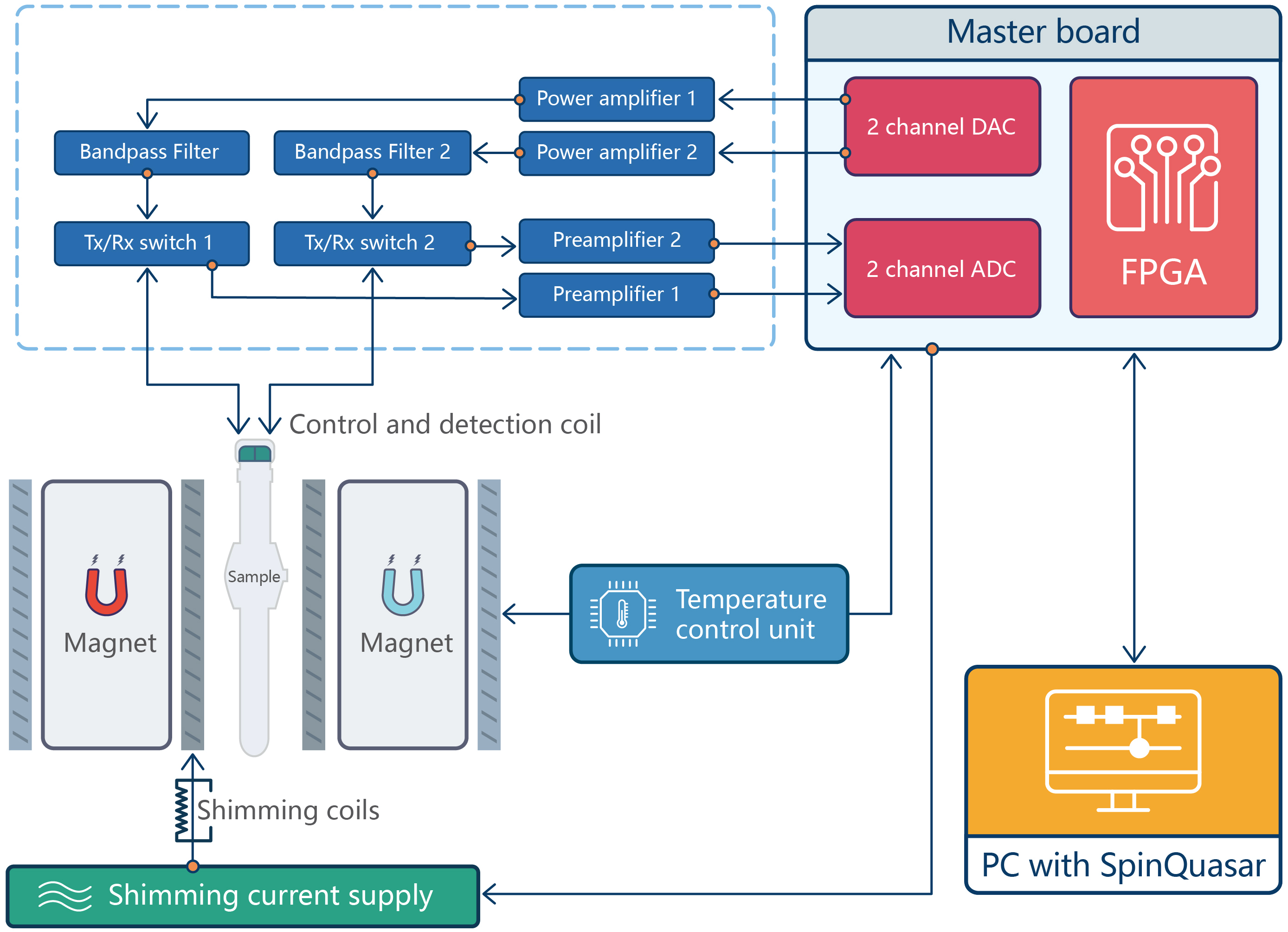}
\caption{The overview of the schematic diagram of Gemini system and the interface \textsc{SpinQuasar}. The master board equipped with an FPGA, provides the control logic of Gemini. \textsc{SpinQuasar} communicates with FPGA through USB so that the user can access Gemini. The magnets, together with the temperature control unit and the field shimming system provide a stable static homogeneous magnetic. The RF module provides the function required to control and measure the qubits.}
\label{hardware}
\end{figure*}

The overall schematic diagram is shown in Fig.~\ref{hardware}.
Gemini is composed of  a PC with \textsc{SpinQuasar}, a control system on the master board, a radio frequency (RF) system, a temperature control module, a pair of two permanent magnets, a field shimming system, and  a tube of sample. 

The PC with \textsc{SpinQuasar} and the master board together realize the algorithms and interfaces to all the functions. 
The magnets provide static stable magnetic field. The field shimming system and temperature control system together make the field stable and homogeneous enough  for nuclear magnetic resonance as well as quantum computing.  The RF system provides generation, modulation, amplification, transmission, detection and reception of the RF pulses so that we can control and measure the quantum system.

\begin{figure}[!ht]
\includegraphics[scale=0.08]{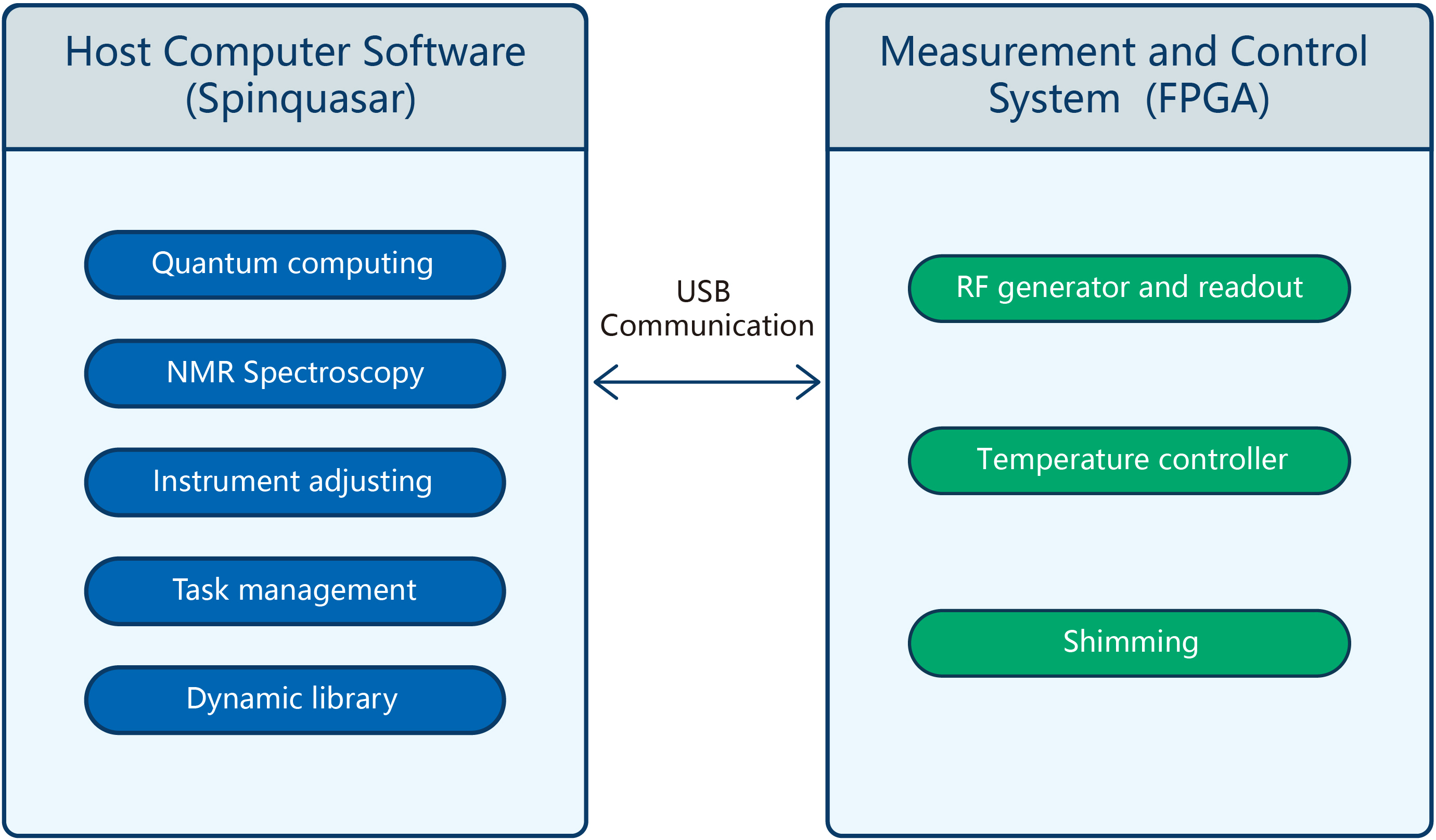}
\caption{Software structure. The software could be divided into two parts: one (the left block) we called \textsc{SpinQuasar} provides an interface for users with access to all the functions of Gemini, the other (the right block) realizes the algorithms and controls required for Gemini to function properly. These two parts communicate with each other through USB.}
\label{intcontrol}
\end{figure}

The PC with \textsc{SpinQuasar} and the master board altogether realized the software part. The modules are shown in \cref{intcontrol}. The software \textsc{SpinQuasar} provide an interface for a user to communicate with the quantum computer Gemini. The master board, of which the core device is an FPGA, realizes all the algorithms to control the pulses (hence control the quantum state), the temperature and shimming of the field (hence generate a stable homogeneous field). \textsc{SpinQuasar} and the master board communicate with each other through USB.

\subsection{SpinQuasar}

The left half of \cref{intcontrol} shows the structure of \textsc{SpinQuasar}. It is composed of five modules: the quantum computing module, the NMR spectroscopy module, the instrument calibration module, the task management module and the dynamic library module. These modules can be easily accessed from the homepage of \textsc{SpinQuasar}, as shown in the right column of \cref{photo}.

The quantum computing module provides an interface of a two-qubit quantum computer and will be discussed later in Sec. III.

The NMR spectroscopy module provides an interface of direct accessing to the $^1$H and $^{31}$P nuclear magnetic resonance signal of our sample. It provides the direct control of the pulse parameters on the two nuclei. Also, it provides both the free induction decay (FID) signals and the spectra after fast Fourier transform (FFT). This interface provides a good demonstration of a modern FFT based NMR spectrometer.

The instrument calibration module provides an interface for users to calibrate the parameters of the spectrometer, such as field shimming, phase calibration, and the temperature control for the system.

Our quantum computer also supports cloud quantum computing. For cloud quantum computing, the tasks are  managed by the task management module. Also, to support more complicated control, such as the variational quantum eigensolver (VQE) which requires adjusting the parameters of the pulses, we provide the APIs for programmable control, and embedded these into the dynamic library. 

\subsection{Master board}

The master board integrates the digital parts of the hardware, including an FPGA, an analog-digital converter (ADC) and a digital-analog converter (DAC). The digital parts, as shown in  the right block  of \cref{intcontrol},  altogether realize the algorithms required to generate the RF pulse, measure the readout signal, control the temperature and shimming. These algorithms will be described further in the introduction of each module. The ADC converts the readout signal from the RF part as measurement, while the DAC generate the initial RF signal for state manipulation

\subsection{Magnets}

The permanent magnets provide a stable static homogeneous magnetic field, which split the nuclei with spin-half into two energy levels and therefore become a qubit. The permanent magnets are two NdFeB plates.  The field generated is $\sim$1 Tesla. The field near the center of the two magnets is roughly homogeneous: the homogeneity generated by such magnets can reach a level of $\sim$20 ppm. Compared with modern commercial NMR spectrometers, of which the magnet is generated by the superconducting coil which requires  a large cryogenic storage dewar and regular refilling of liquid helium and nitrogen, the magnets of Gemini works under room temperature, hence maintenance free and portable. The disadvantages are that the magnitude of the magnets could only reach about 2 T, and the magnetic field is highly sensitive to the temperature of the magnets themselves. 

\subsection{Sample}

\begin{figure}[!ht]
\includegraphics[width=8cm]{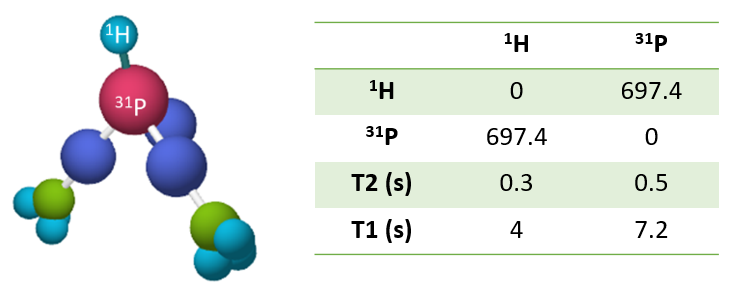}
\caption{The molecule structure (left) and its parameter table (right). The J coupling between the $^1$H and $^{31}$P nuclear spins is 697.4 Hz. The control pulses are on resonance with $^1$H and $^{31}$P spins and thus their frequency offsets are both 0 Hz. The spin Hamiltonian is $H_0=\pi J \sigma_z^{H}\sigma_z^{P}/2$, where $J=697.4$Hz.}
\label{molecule}
\end{figure}

The sample we used is Dimethylphosphite ((CH$_3$O)$_2$PH) molecules . The $^{31}$P and $^{1}$H atom are connected directly and provide a two-qubit quantum processor. Both $^{31}$P and $^{1}$H nuclei have a $1/2$-spin, and therefore have two energy levels. The Lamor frequency of  $^{31}$P and $^{1}$H in 1 T magnetic field are 17.2 MHz and 42.6 MHz, respectively. The structure and the parameters of the sample are listed in (Fig. \ref{molecule}).

\subsection{RF pulse generation}

The states of the nuclei could be manipulated by irradiating electro-magnetic waves (pulses) with frequencies close to there Larmor frequency (the physics behind this will be described later). Since the Larmor frequencies of the two nuclei lie in the range of RF range, an RF system is designed and manufactured to operate the quantum state and realize the quantum gates. 

\begin{figure}[!ht]
\includegraphics[width=8cm]{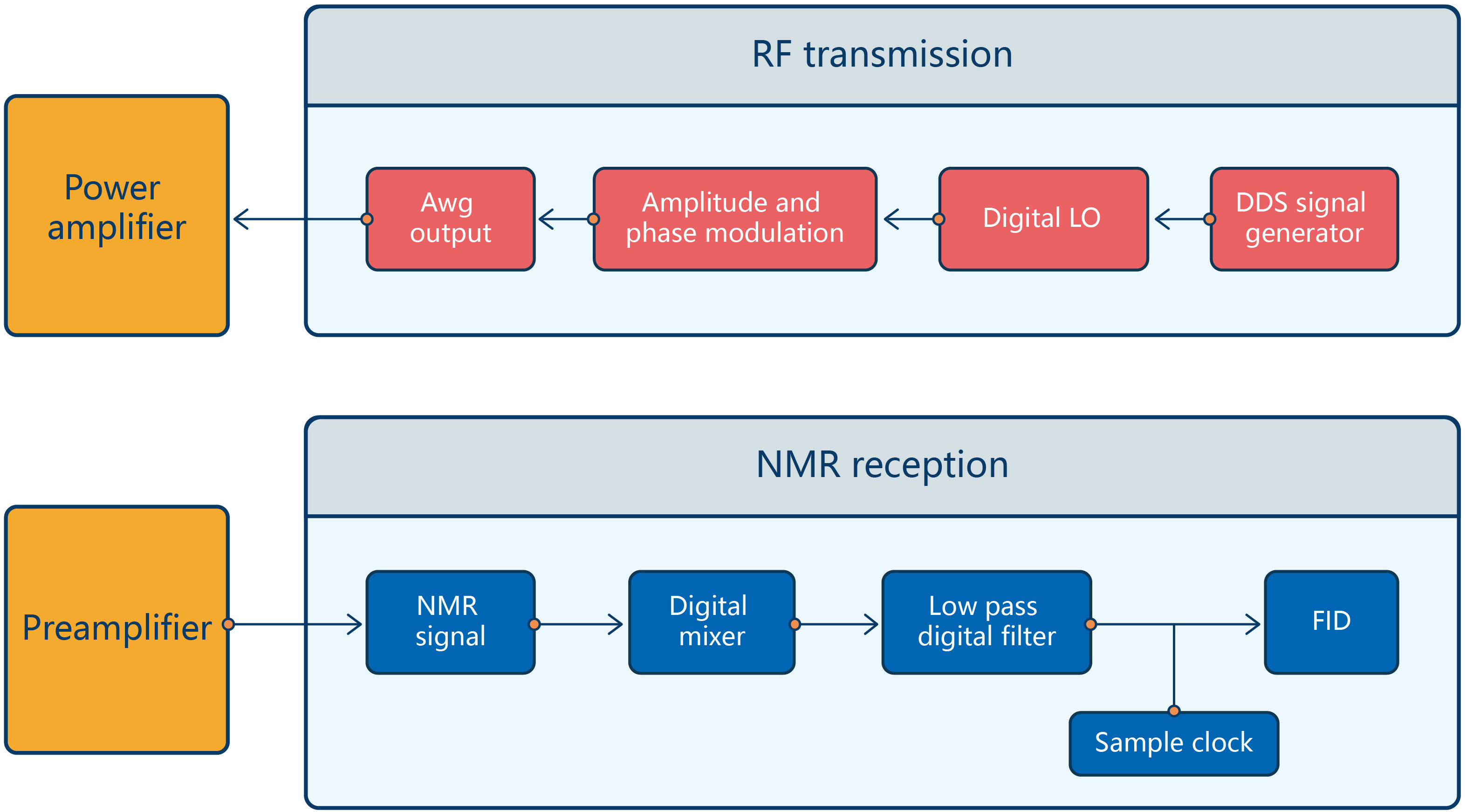}
\caption{RF pulse generation and signal readout. The upper half shows how an arbitrary wave is generated. After the wave is generated, it is power amplified and sent to the control coil (see \cref{hardware}) so that the quantum state can be manipulated. After the RF pulse irradiation, the electro-magnetic signal induced during spin relaxation is picked up by the coil, and then is sent to the preamplifier. After the signal is amplified, it is sent to the ADC and processed by the master board.}
\label{rf}
\end{figure}

\subsection{Temprature control}

The field generated by the permanent magnets is highly sensitive to the temperature of the permanent magnets themselves. Therefore, a temperature control system is required to guarantee that the field does not drift following the room temperature. 
\begin{figure}[!ht]
\includegraphics[width=8cm]{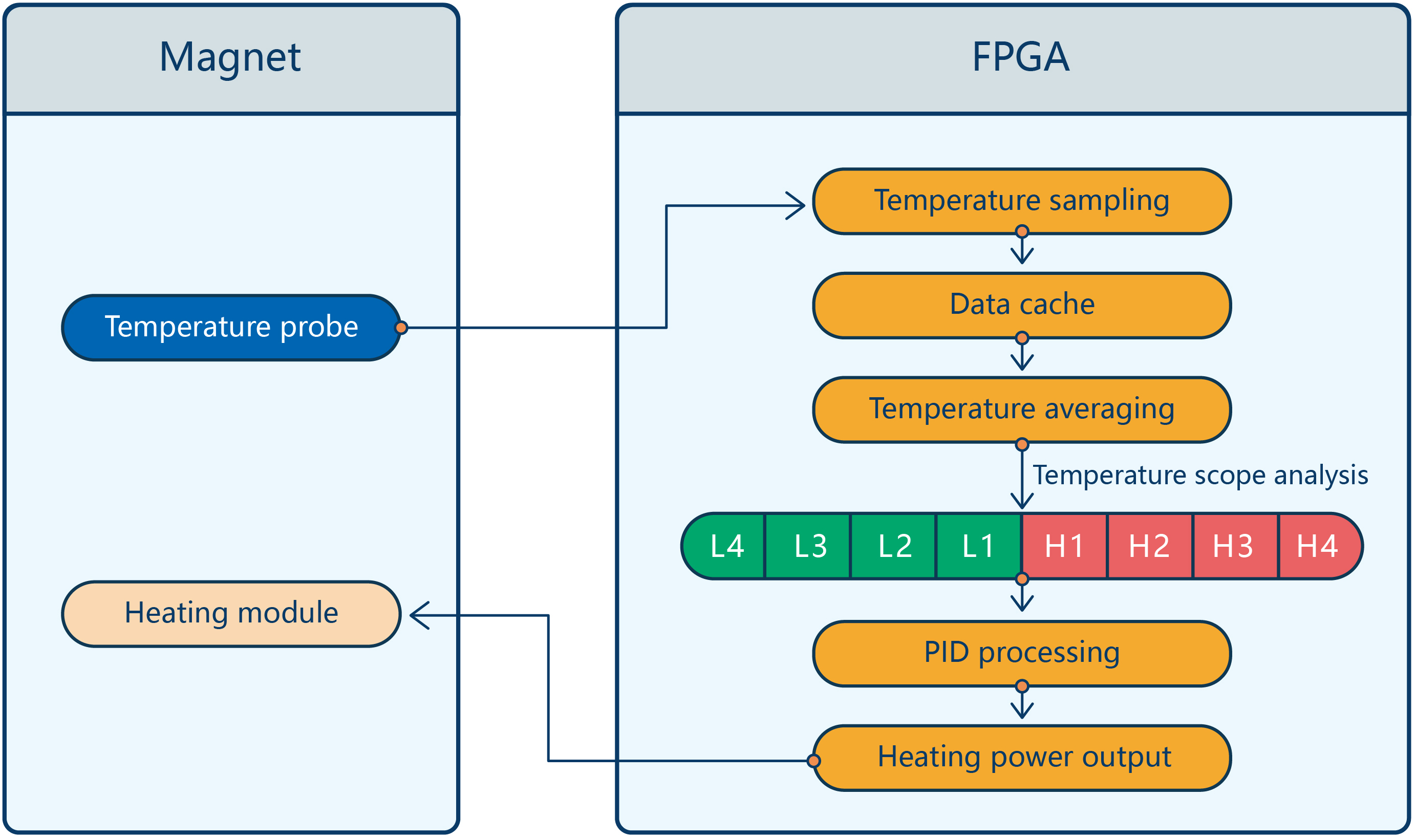}
\caption{Temperature control. The temperature control is realized by a famous feedback algorithm: the PID algorithm. The temperature probe picks the temperature signal and send it to the FPGA. The FPGA decides what to do next according to the temperature signal and then controls the power of the heating module accordingly.}
\label{tempcont}
\end{figure}

\subsection{Field Shimming}
The homogeneity of the static magnetic field generated by the permanent magnets is $\sim$20 ppm, which is too large. To compensate this inhomogeneity, we designed a field shimming system to reduce the homogeneity to less than $\sim 1$ ppm. The best homogeneity could be reached to $\sim 0.3$ ppm. As a comparison, the homogeneity of a commercial nuclear magnetic spectrometer with a superconducting magnet is $\sim 0.01$ ppm.

\begin{figure}[!ht]
\includegraphics[width=8cm]{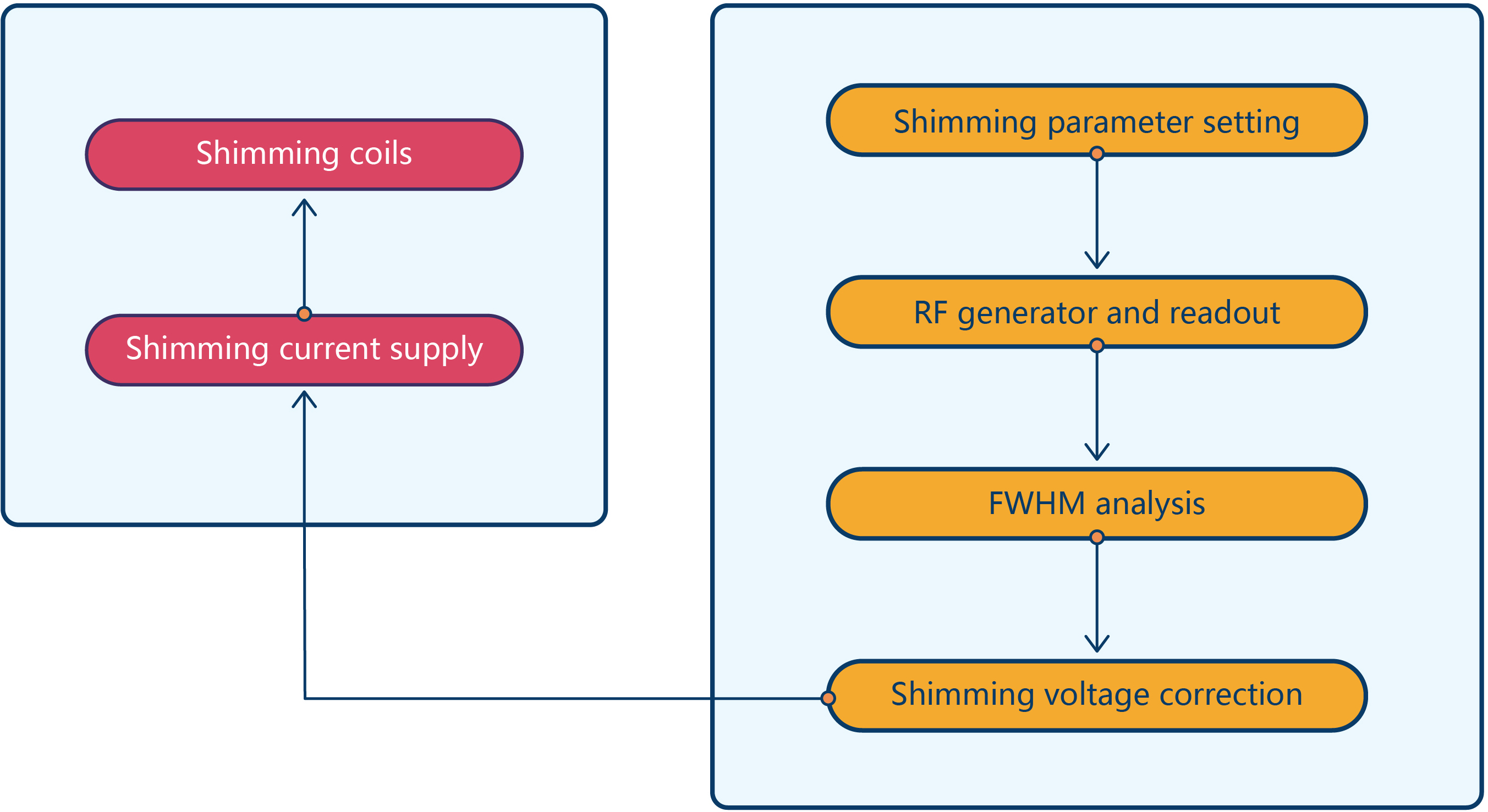}
\caption{Shimming. Currently, the field shimming is accomplished by measuring the FWHM of the hydrogen signal. It is also a feedback algorithms by reading the FWHM information of the hydrogen signal and adjust the current of the shimming coils. }
\label{shim}
\end{figure}

\section{Quantum computation}
\subsection{The spin system}
 Gemini contains two qubits which are the two connected $^{31}$P and $^{1}$H nuclear spins in Dimethylphosphite ((CH$_4$O)$_2$PH) molecules (Fig. \ref{molecule}). The molecules are placed in the center of the parallel permanent magnets. The $^{31}$P and $^{1}$H larmor frequencies are 17.2 MHz and 42.6 MHz, respectively. The $^{31}$P spin has a T$_1$ and T$_2$ of 7.2 s and 0.5 s, respectively. The  $^{1}$H  spin has a T$_1$ and T$_2$ of 4 s and 0.3 s, respectively. The J coupling between the two spins is  697.4 Hz. The control pulses are on resonance with $^1$H and $^{31}$P spins and thus their frequency offsets are both 0 Hz. The spin Hamiltonian in the rotating frame is 
\begin{align} 
H_0=2\pi J I_z^{H}I_z^{P}=\frac{\pi}{2} J \sigma_z^{H}\sigma_z^{P},
\end{align}
 where $J=697.4$ Hz.

\subsection{The gate set}
Single-qubit 90 degree rotation gates can be realized using square pulses of 20us and 10us for $^{31}$P and $^{1}$H, respectively. The hardware-level pulse design and engineering are available in later versions of Gemini which provide an arbitrary waveform generation function to users. In the current paper, all quantum gates are realized using square pulses which are resonant with $^{1}$H or $^{31}$P and combined with free evolution. The available quantum gates contain single-qubit and two-qubit gates. The single-qubit gates are as follows:
\begin{align}
&X=\sigma_{x}=\begin{pmatrix}
0 & 1\\
1 & 0
\end{pmatrix},
Y=\sigma_{y}=\begin{pmatrix}
0 & -i\\
i & 0
\end{pmatrix},
Z=\sigma_{y}=\begin{pmatrix}
1 & 0\\
0 & -1
\end{pmatrix},\nonumber\\
&X90=e^{-i\frac{\pi}{4}\sigma_x},Y90=e^{-i\frac{\pi}{4}\sigma_y},Z90=e^{-i\frac{\pi}{4}\sigma_z},\nonumber\\
&Rx=e^{-i\frac{\alpha}{2}\sigma_x},Ry=e^{-i\frac{\beta}{2}\sigma_y},Rz=e^{-i\frac{\gamma}{2}\sigma_z},\nonumber\\
&H=\frac{1}{\sqrt{2}}\begin{pmatrix}
1 & 1\\
1 & -1
\end{pmatrix},
I=\begin{pmatrix}
1 & 0\\
0 & 1
\end{pmatrix}.
\end{align} 
Here, $\alpha$, $\beta$ and $\gamma$ are the rotation angles defined by users. The two-qubit gates are as follows
\begin{align}
&\mathrm{CX}=\begin{pmatrix}
1 & 0 & 0 & 0\\
0 & 1 & 0 & 0\\
0 & 0 & 0 & 1\\
0 & 0 & 1 & 0
\end{pmatrix},
\mathrm{CY}=\begin{pmatrix}
1 & 0 & 0 & 0\\
0 & 1 & 0 & 0\\
0 & 0 & 0 & -i\\
0 & 0 & i & 0
\end{pmatrix},\nonumber\\
&\mathrm{CZ}=\begin{pmatrix}
1 & 0 & 0 & 0\\
0 & 1 & 0 & 0\\
0 & 0 & 1 & 0\\
0 & 0 & 0 & -1
\end{pmatrix},
\mathrm{delay}=e^{-itH_0}=e^{-it\frac{\pi}{2}J\sigma_z^{H}\sigma_z^{P}},
\end{align} 
where $\mathrm{CX}$ gate is the famous control NOT ($\mathrm{CNOT}$) gate.
Here, the delay gate is a free evolution gate with the duration $t$ defined by users. It should be noted that when $t$ is large noise plays a non-negligible role and the action of this gate is not ideal as the form in the above equation. The single-qubit gate fidelity is estimated to be $\sim$0.99 and the two-qubit gate fidelity is estimated to be $\sim$0.98.

\subsection{The pseudo-pure state}
The initial state of the two-qubit system is prepared to be a pseudo-pure state (PPS) \cite{Corypps}. The thermal equilibrium state of a liquid-state NMR system is subject to Boltzmann distribution and at room temperature can be expressed as follow:
\begin{align}
\rho_{\mathrm{eq}}=\frac{e^{-H_s/k_BT}}{\mathrm{Tr}(e^{-H_s/k_BT})}\approx\frac{1}{2^n}I^{\otimes n}+\Sigma^n_{k=1}\frac{1}{2}\epsilon_k\sigma_z^k.
\end{align}
Here, $H_s$ is the spin Hamiltonian in the lab frame, and $n$ is the number of qubits. The part $\Sigma^n_{k=1}\frac{1}{2}\epsilon_k\sigma_z^k$ gives NMR signals. At room temperature $\epsilon\sim e^{-5}$  is small and thus this is a highly mixed state. To implement quantum computation, researchers  \cite{Corypps} proposed to use PPS as the initial state which has the following form,
\begin{align}
\rho_{\mathrm{pps}}=\frac{1-\eta}{2^n}I^{\otimes n}+\eta|\psi\rangle\langle\psi|.
\end{align}
$|\psi\rangle$ is a pure state. The PPS above has the same unitary dynamics and observable effects as the pure state $|\psi\rangle$ except for the factor $\eta$. PPS is widely used in NMR quantum computation. 

\begin{figure}[!ht]
\includegraphics[width=8cm]{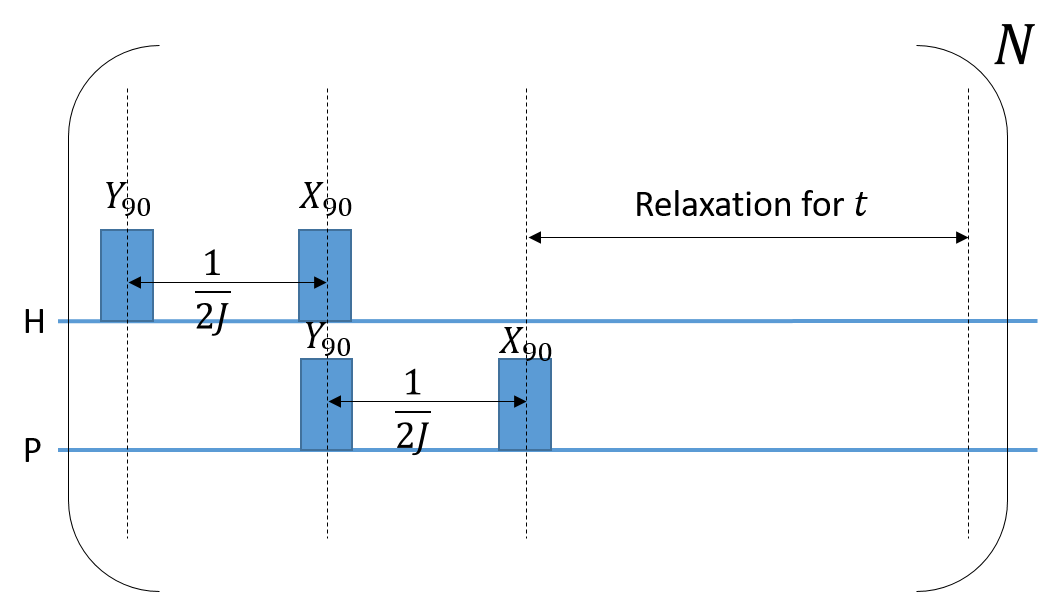}
\caption{The pulse sequence for pseudo-pure state preparation. The first four pulses realize a basis permutation gate. After it is a long delay within which the natural relaxation takes effect. By properly choosing the repetition number, $N$, and the duration of the delay, $t$, the system can be steered to the pseudo-pure state $|00\rangle$ from the thermal equilibrium state.  }
\label{pps}
\end{figure}

Gemini utilizes the relaxation method in Ref. \cite{pps} to prepare the two-qubit PPS starting from the thermal equilibrium state. As shown in Fig. \ref{pps}, the first four pulses realize a basis permutation gate which can be expressed as
\begin{align}
&U_{\mathrm{permute}}=\begin{pmatrix}
-i & 0 & 0 & 0\\
0 & 0 & 0 & -i\\
0 & -1 & 0 & 0\\
0 & 0 & 1 & 0
\end{pmatrix}.
\end{align}
$U_{\mathrm{permute}}$ permutes the basis $|01\rangle$, $|10\rangle$ and $|11\rangle$ and leaves $|00\rangle$ unchanged upon a phase. The relaxation method in Ref.~\cite{pps} combines $U_{\mathrm{permute}}$ and a delay after it during which T1 relaxation takes effect. By properly choosing the number of the repetitions of this combination and the delay time $t$, the system can reach a state whose dominantly occupied basis is $|00\rangle$ and the other three base have the same but smaller probability. This obtained state is a PPS and can be used as the initial state $|00\rangle$ in NMR quantum computing. 
\subsection{Density matrix reconstruction}
\begin{figure}[!ht]
\includegraphics[width=8cm]{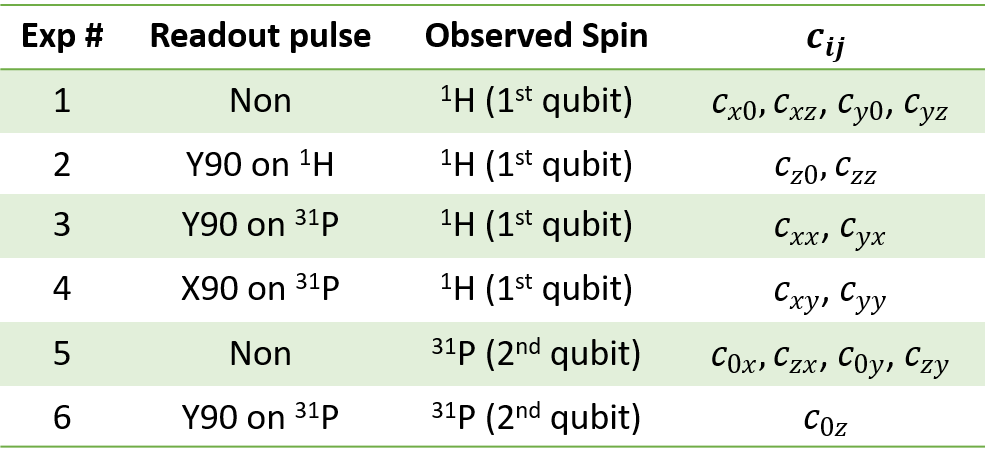}
\caption{The readout pulses, observed spins and the obtained $c_{ij}$ of the six experiments needed to reconstruct a density matrix in the form of Eq. (\ref{qstdecomp}).}
\label{qst}
\end{figure}

Gemini implements quantum state tomography~\cite{QST} to reconstruct the density matrix of the quantum state after a certain gate sequence is applied. Any two-qubit density matrix can be expressed in the following way,
\begin{align}
\rho&=\frac{1}{4}I^{\otimes 2}+\frac{1}{4}\sum_{i,j}c_{ij}\sigma_i\sigma_j,\nonumber\\i(j)&=x,y,z,0, \mathrm{but}\; (i,j)\neq (0,0).\label{qstdecomp}
\end{align}
Here $\sigma_0=I$ is the 2$\times$2 identity matrix. To reconstruct a density matrix, one need to measure all the $c_{ij}$ which are $c_{ij}=\mathrm{Tr}(\rho\sigma_i\sigma_j)$, in other words, the expectation values of the Pauli matrices $\sigma_i\sigma_j$. There are total 15 of $\sigma_i\sigma_j$. But only \{$\sigma_xI$, $\sigma_x\sigma_z$,$\sigma_yI$, $\sigma_y\sigma_z$,$I\sigma_x$, $\sigma_z\sigma_x$,$I\sigma_y$, $\sigma_z\sigma_y$\} are observables in NMR. Additional readout pulses are needed to transform the unobservable components to be observable. For example, by applying a readout pulse $Y90$ prior to measurement, $c_{z0}$ can be obtained,  $c_{z0}=\mathrm{Tr}(Y90\rho Y90^{\dagger}\sigma_xI)$. In Gemini, the reconstruction is realized by repeating a experiment six times, each time with a different readout pulse and  observing either  $^{31}$P or $^{1}$H. The readout pulses and $c_{ij}$ obtained in each of the six repetitions are listed in Fig. \ref{qst}. The reconstructed PPS has a fidelity of higher than 0.99.

\subsection{Software interface}

\begin{figure}[!ht]
\includegraphics[width=8cm]{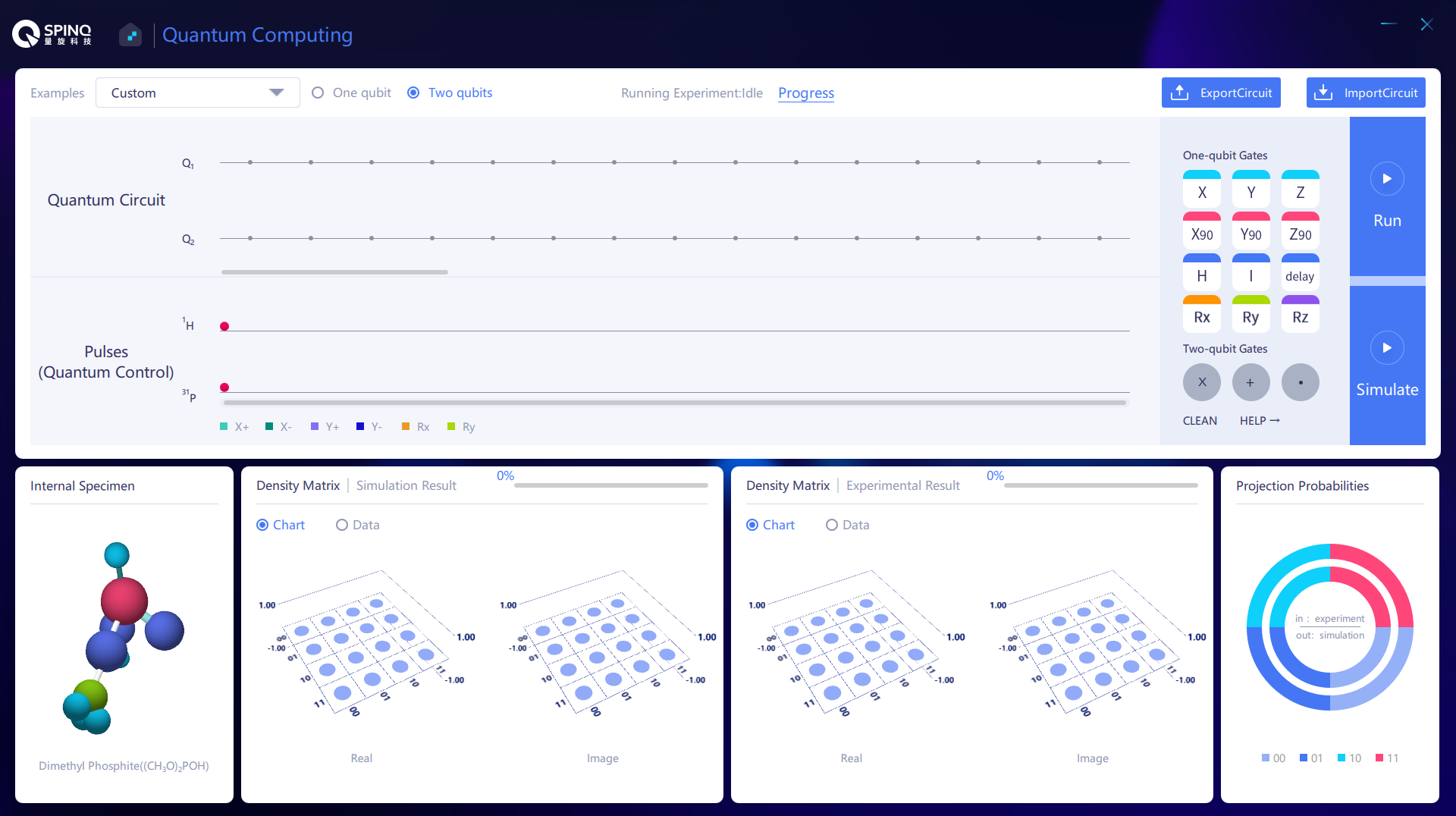}
\caption{The \textsc{SpinQuasar} interface for quantum computing. There is a quantum circuit composer where users can drag and drop the supported quantum gates to construct a desired circuit. The corresponding pulse sequence is shown below the quantum circuit. There are two buttons, 'Run' and 'Simulate', for activation of the experiment and the simulation, respectively. The density matrices from the experiment and the simulation are shown in the bottom half of the interface.}
\label{qcint}
\end{figure}

\begin{figure}[!ht]
\includegraphics[width=8cm]{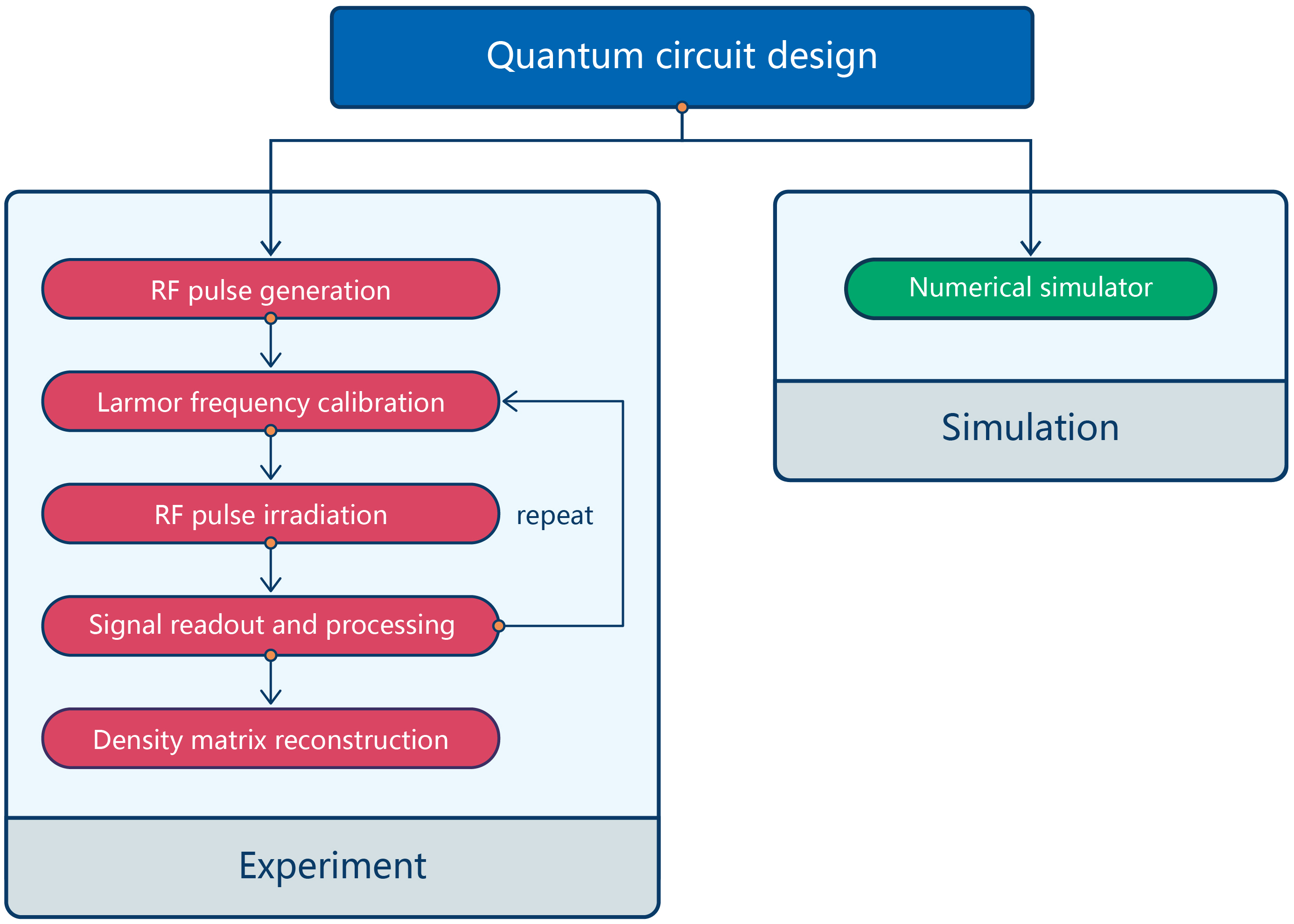}
\caption{Realization of the quantum computing system. It consists of the experiment on the two-qubit processor and the numerical simulation.}
\label{qc}
\end{figure}

The user can use the quantum computing interface of \textsc{SpinQuasar} to access the quantum computing function of Gemini (\cref{qcint}).
The structure and flow-chart of quantum computing is shown in Fig. \ref{qc}. This system wraps up the calibrated pulses into the quantum gates aforementioned. Users can drag the supported gates into the circuits and press Run, the two-qubit quantum computer will start running. The final result will be shown in the form of density matrix which is reconstructed in the way discussed in the last section. There is also a noiseless simulator embedded in this system so that one can easily compare the experimental results with theoretical results. 
 

\section{Application: Measurement of geometric phase of mixed state}
Gemini provides demonstrations of $>10$ famous quantum algorithms, such as Deutsch-Jozsa algorithm \cite{deutsch1992rapid}, Grover search~\cite{grover1996fast}, and HHL algorithm~\cite{harrow2009quantum}.
In this paper, we provide two more advanced examples that demonstrate Gemini's ability on running quantum algorithms. In this section, we will demonstrate the measurement of the geometric phase of mixed states.

\subsection{Theory}
Geometric phase \cite{geometricReview} is a very important concept in quantum mechanics. It is a type of phase gained by a system that is determined by the geometry of the system's evolution path. The most famous geometric phase is berry phase \cite{Berryphase} which is associated with cyclic adiabatic evolution. Here we use a spin half system as an example. A spin in a magnetic field is aligned with the field and is in the state $|0\rangle$. If the field direction changes slowly and  the adiabatic conditions are satisfied, the spin direction changes also adiabatically and is always along the field direction. When the magnetic field returns to its initial direction, the spin returns to its initial direction as well. However, the spin state gains a global phase and is $e^{i(\alpha+\beta)}|0\rangle$, where $\alpha$ is the dynamic phase and $\beta$ is the Berry phase. $\alpha$ and  $\beta$ have expressions as follows:
\begin{align}
\alpha&=-\frac{1}{\hbar}\int_0^\tau E(t)\mathrm{d}t,\\
\beta&=-\frac{1}{2}\Omega.
\end{align}
$E(t)$ is the energy of $|0\rangle$ at the time $t$ and is determined by the instant Hamiltonian. $\Omega$ is the solid angle enclosed by the path. If the initial state is in $|1\rangle$, and the magnetic field changes along the same path, then the geometric phase gained by the spin is $\Omega/2$. This is because the spin is opposite in this case and hence its path encloses a solid angle of $-\Omega$.

Berry phase is discussed above in the context of adiabatic evolution. Researchers have proved that adiabatic evolution is not a necessary condition for geometric phase~\cite{nonadiabaticphase}. Geometric phase stays the same as long as the geometry of the evolution path stays the same, and is not affected by the Hamiltonian that drives this evolution.

Geometric phase is believed to be robust to local noise and fluctuations of Hamiltonian parameters because of its connection with the path geometry. Therefore, geometric quantum computation is proposed as a candidate for fault-tolerant quantum computation~\cite{geometricNMR,Holonomic,GeometricIon,GeometricWang,HolonomicLidar,fault-TolerantHolonomic,PhysRevLett.87.097901,Sj_qvist_2012,NonHoloXU,NonHoloExpFeng,NonHoloExpDuan}. Geometric phase in noisy environments is also studied. When the environment is noisy, quantum systems are always in mixed states due to the interaction with the environment. The work in Ref. \cite{Geomixed} provides a definition for the geometric phase of a mixed state: It is the phase shift of the interference oscillations in interferometry gained by the mixed state after a unitary evolution. The unitary evolution must satisfy the parallel transport requirement \cite{geometricReview,Geomixed}: The state at any instant is in-phase with the state after an infinitesimal time. It can be proved that the dynamical phase is 0 if the parallel transport requirement can be satisfied. After such a unitary evolution, each eigen state of the density matrix of the initial mixed state gains a phase denoted as $\gamma_n$, and has the interference visibility $\nu_n$. The geometric phase $\gamma$ of the mixed state and its interference visibility $\nu$ satisfy the following equation:
\begin{align}
\nu e^{i\gamma}=\Sigma_n p_n\nu_n e^{i\gamma_n}.
\label{geodef}
\end{align}
Here, $p_n$ is the eigen value of the $n$th eigen state of the density matrix. 

\subsection{Experimental protocol}
\begin{figure}
\includegraphics[width=5cm]{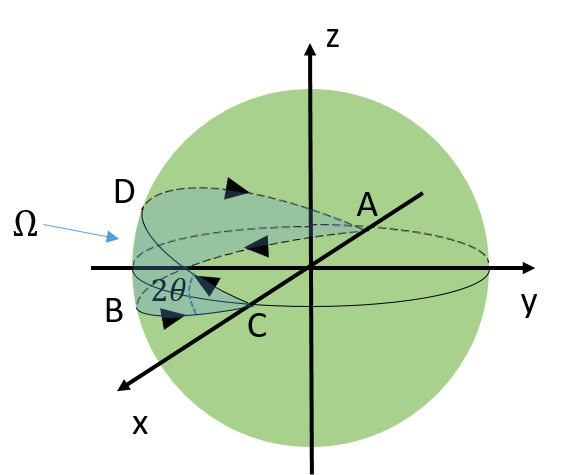}
\caption{The unitary evolution path for the mixed state. The state vector of the initial mixed state is prepared to be along -x axis pointing to A. It evolves along the path A-B-C-D-A to return back to point to A. $\Omega$ is the solid angle enclosed by this path. $\theta=\Omega/4$ is the angle between the x-y plane and either of the half paths (A-B-C or C-D-A).}
\label{bloch}
\end{figure}

We adapt the protocol used in Ref. \cite{geometric} to measure the geometric phase in mixed states as defined in Eq. \ref{geodef}. A two-qubit system is used in this protocol. The first qubit is an ancilla qubit and the second qubit is in the mixed states whose geometric phase is to be measured. The mixed state is a mix of $|+\rangle=\sqrt{2}(|0\rangle+|1\rangle)/2$ and $|-\rangle=\sqrt{2}(|0\rangle-|1\rangle)/2$. The initial mixed state is:
\begin{align}
\rho(0)=\frac{1}{2}(I+\vec{\bm{r}}\cdot \vec{\bm{\sigma}})=\frac{1}{2}(I-r\sigma_x).\label{initial}
\end{align}
Here $\vec{\bm{r}}$ is the Bloch vector, and $r$ is its length that corresponds
to the purity of the state. If $r$ = 1, the state is a pure state which is $|-\rangle$. If $r$ = 0, the state is totally mixed. $|-\rangle$ and $|+\rangle$ are the
two eigen states of the above density matrix with eigen values of $(1+r)/2$ and $(1-r)/2$. Here we steer the state along the path (A-B-C-D-A) which encloses a solid angle $\Omega$ as shown in Fig. \ref{bloch}. Because the path is made up of geodesic curves, the parallel transport requirement can be satisfied and thus the dynamical phase is zero. The $|-\rangle$ and $|+\rangle$ states gain geometric phases of $-\Omega/2$ and $\Omega/2$, and change to $e^{-i\Omega/2}|-\rangle$ and $e^{i\Omega/2}|+\rangle$, respectively. It can be proved that the interference visibility of the two eigen states are both 1. Then the geometric phase $\gamma$ of the mixed state satisfies
\begin{align}
\nu e^{i\gamma}&=\frac{1}{2}(1+r)e^{-i\frac{\Omega}{2}}+\frac{1}{2}(1-r)e^{i\frac{\Omega}{2}}=\cos\frac{\Omega}{2}-ir\sin\frac{\Omega}{2}\\
\gamma&=-\tan^{-1}(r\tan\frac{\Omega}{2})\label{mixedphase}
\end{align}
In order to measure $\gamma$, the ancilla qubit is prepared in the state $\sqrt{2}(|0\rangle_a+|1\rangle_a)/2$, which has the density matrix $(I+\sigma_x^a)/2$. We control the two-qubit system so that when the ancilla qubit is in $|1\rangle_a$, the mixed state undergoes the unitary evolution and when the ancilla qubit is in $|0\rangle_a$ nothing happens. The phases gained by $|-\rangle$ and $|+\rangle$ in the mixed state (which are $\mp\Omega/2$) are passed to the ancilla qubit. Thus after the controlled evolution, the state of the ancilla qubit is $\sqrt{2}(|0\rangle_a+e^{\mp i\Omega/2}|1\rangle_a)/2$. The weighted average phase gained
by the ancilla qubit has the form of Eq. (\ref{mixedphase}).

Next, we discuss how to prepare a mixed state in the form of Eq. (\ref{initial})  from the initial PPS state $|00\rangle$. The most used method in NMR to prepare such a mixed state is to use a pulsed gradient field, which can dephase the spin polarization in the x-y plane fast. However, there is no pulsed gradient field in Gemini. Considering the time scale of dephasing caused by the static field inhomogeneity
as well as T$_2$ is much smaller than T$_{1}$, we exploit the natural dephasing to remove the unwanted polarization. To prepare a state in Eq. \ref{initial}, first the state $(I+r\sigma_z)/2$ is prepared from $|0\rangle$:
\begin{align}
|0\rangle=\frac{1}{2}(I+\sigma_z)\xrightarrow{R_x(\cos^{-1}r)}\frac{1}{2}(I+r\sigma_z-\sqrt{1-r^2}\sigma_y).
\end{align}
The $-\sqrt{1-r^2}\sigma_y$ part in the above equation can be removed using natural dephasing and we can get $(I+r\sigma_z)/2$. Then, rotate $(I+r\sigma_z)/2$ about $y$ axis by $-\pi/2$, we get $(I-r\sigma_x)/2$.

\begin{figure*}[!ht]
\includegraphics[scale=0.3]{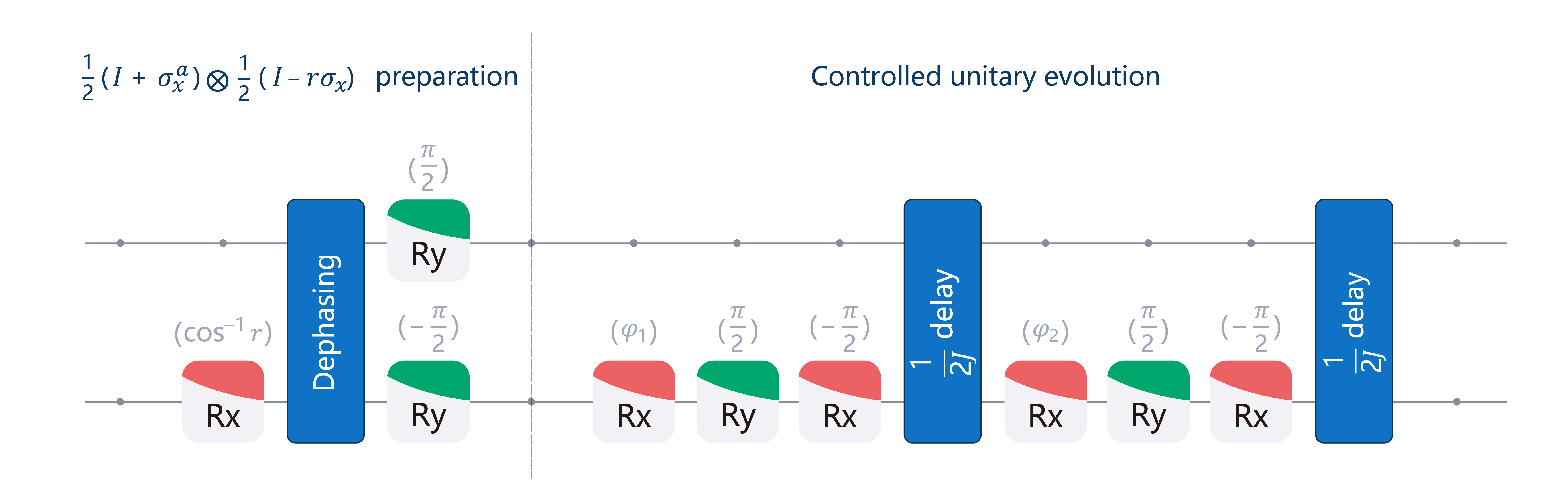}
\caption{The experimental circuit for the measurement of the geometric phase for mixed states.}
\label{geocircuit}
\end{figure*}

The $|1\rangle_a$-controlled unitary can be realized using the following sequence.
\begin{align}
R_{x}(-\theta)\to\mathrm{CZ}\to R_{x}(2\theta-\pi)\to\mathrm{CZ}\label{CU}
\end{align}
$\theta$ is the angle between either of the two half paths and the x-y plane, $\theta=\Omega/4$. $R_{x}(-\theta)$ operation rotates the first half of the path to the x$<$0 half of the x-y plane. CZ is the controlled-Z gate. When the ancilla qubit is in $|0\rangle_a$, CZ does nothing; when the ancilla qubit is in $|1\rangle_a$, CZ rotates the mixed state about z axis by $\pi$ counterclockwise, which means the mixed state evolves along the first half of the path. $R_{x}(2\theta-\pi)$ rotates the second half of the path to the x$>$0 half of the x-y plane. The CZ after it realizes the evolution of the mixed state along the second half of the path when the ancilla qubit is $|1\rangle_a$. In this way, the mixed state undergoes a closed path evolution conditional on the $|1\rangle_a$ state of the ancilla qubit. CZ can be further decomposed as
\begin{align}
R_x(\frac{\pi}{2})\to R_y(\frac{\pi}{2})\to R_{x}(-\frac{\pi}{2})\to\frac{1}{2J},
\end{align}
here $1/2J$ refers to the free evolution for a duration of $1/2J$ under the J coupling between the two qubits. The $R_x(\pi/2)$ gate in CZ can be combined with the $R_{x}(-\theta)$  and $R_{x}(2\theta-\pi)$ operations in Eq. \ref{CU} and simplified. After this simplification, the quantum circuit is shown in Fig. \ref{geocircuit}. The first qubit is the ancilla qubit. $\phi_1=\pi/2-\theta$, $\phi_2=2\theta-\pi/2$. The geometric phase $\gamma$ can be measured by measuring the phase change of the ancilla qubit after implementing the circuit in Fig. \ref{geocircuit}.

\subsection{Results}

\begin{figure}[!ht]
\includegraphics[width=8cm]{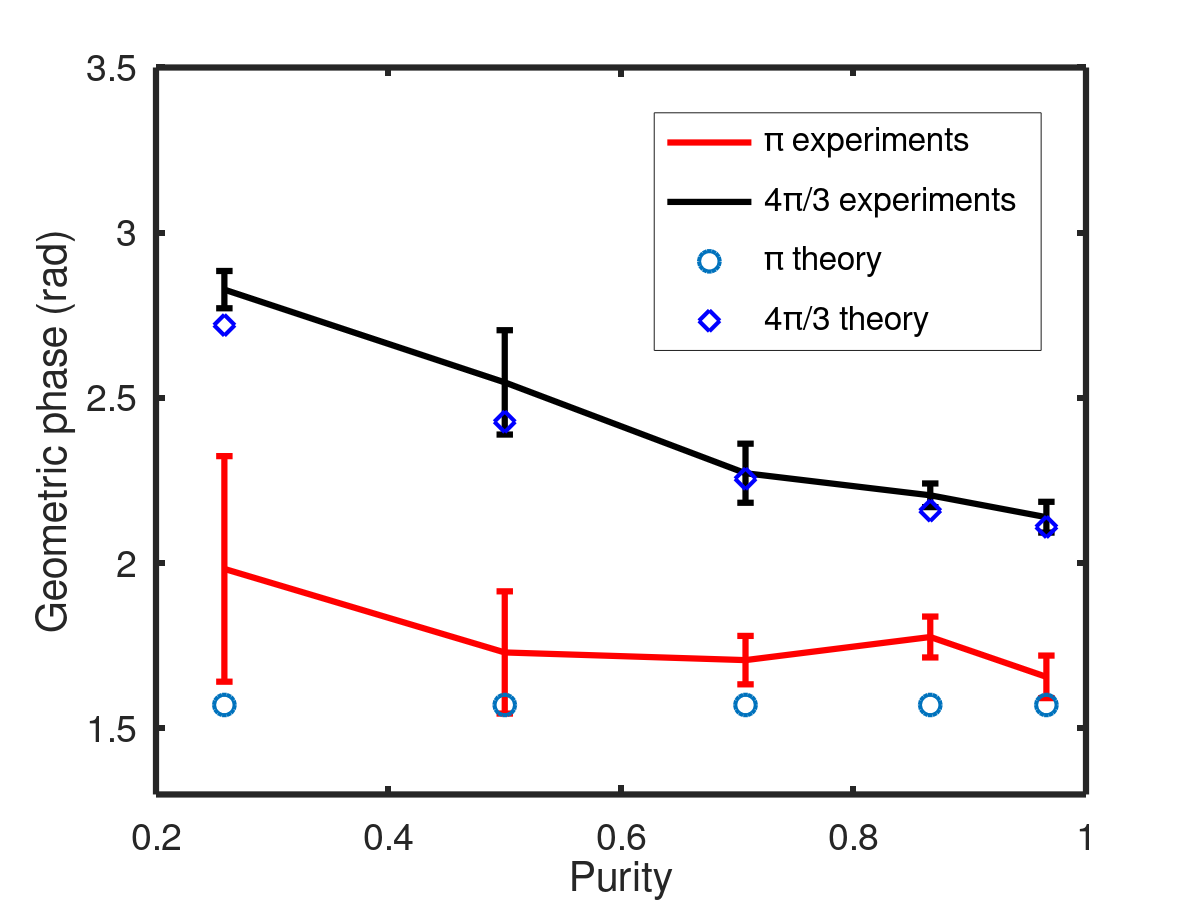}
\caption{The theoretical and experimental results of the mixed state geometric phase in two situations: $\Omega= 180^{\circ}$ and $\Omega= 240^{\circ}$. For each situation, five purity values were chosen for the initial mixed state, $r$=[0.26, 0.50, 0.71, 0.87, 0.97]. For each experimental data point, the experiments were repeated for 5 times and the mean value of the geometric phase was calculated. The error bar indicates the standard deviation of the geometric phase in the 5 repetitions. The theoretical values of the geometric phase are also shown as circles ($\Omega= 180^{\circ}$) and diamonds ($\Omega= 240^{\circ}$) in the figure.}
\label{georesult}
\end{figure}

Experiments with $\Omega= 180^{\circ}$ and $\Omega= 240^{\circ}$ are carried out. In each situation, $r$ is chosen to be [0.26, 0.50, 0.71,
0.87, 0.97]. And for each of the $r$ values, the experiment is repeated for five times to get a mean value of the measured phases as the result of 
$\gamma$ (Fig. \ref{georesult}). The main error sources are the non-ideal initial mixed state and RF pulse imperfections, such as finite pulse width. The large fluctuations in the experimental results come from the uncertainty in fitting the NMR spectra. In spite of those errors and imperfections in experiments, the change trend of the geometric phase as a function of the purity and the solid angle of the path can be observed from the results.

\section{Application: variational quantum eigensolver}

In this section, we implement a variational quantum eigensolver for a two-qubit Heisenberg model.

\subsection{Background}

Quantum algorithms such as the Grover search~\cite{grover1996fast}, Shor factorization~\cite{shor1994,shor1999polynomial} and HHL~\cite{harrow2009quantum}, have proven advantages over their best known classical counterparts. However, these algorithms cannot be efficiently implemented on near-term quantum devices due to inevitable physical noises in quantum gates.  Variational quantum algorithms (VQA) ~\cite{cerezo2020variational, Kandala2017, larose2019variational, zeng2020variational, romero2017quantum, cao2020noise, yuan2019theory}, a class of algorithms under the hybrid quantum-classical framework, are more promising to have practical applications on noisy intermediate-scale quantum quantum computers~\cite{preskill2018quantum}. VQA use a parameterized quantum circuit to estimate the cost function $C(\boldsymbol{\theta})$ and update $\boldsymbol{\theta}$ with a classical optimizer. Variational quantum eigensolver (VQE)~\cite{Kandala2017, hempel2018quantum} is a paradigmatic example of VQA that aims to find the ground state and ground state energy of a given Hamiltonian $H$. In this section, we will demonstrate the experimental realization of VQE on Gemini.

\subsection{Algorithm}

In classical computaitional physics (chemistry), we usually estimate the ground state energy of $H$ through variational approaches: parameterize a wave function $|\psi\rangle=|\psi(\bm \theta)\rangle$, update $\boldsymbol{\theta}$ to minimize the expectation value $\langle \psi(\bm \theta)|H|\psi(\bm \theta)\rangle$ until convergence. 
VQE facilitates the above procedure with a quantum computer, the wave function is parameterized with a quantum circuit $U(\bm \theta)$ applied to the initial state $|\bm 0 \rangle = |0\rangle^{\otimes n}$, we optimize $\bm \theta$ to minimize the expectation value,
\begin{equation}
E(\bm\theta)=\langle \bm 0|U^{\dagger}(\bm \theta) H U(\bm \theta) |\bm 0\rangle.
\end{equation} 
The classical optimizer can either be gradient-based methods like SGD, Adam, RMSprop, BFGD, or gradient-free methods like Nelder-Mead, Powell. Hardware-efficient ansatz~\cite{Kandala2017}, unitary coupled clustered ansatz~\cite{lee2018generalized}, and Hamiltonian variational ansatz~\cite{wecker2015progress, wiersema2020exploring} are common choices for $U(\bm \theta)$.
In VQE, the gradient can be directly estimated via the parameter-shift rule~\cite{Mitarai2018, Schuld2019}, i.e., 
\begin{equation}
\frac{\partial E(\bm \theta)}{\partial \theta_i}=(\left<H\right>_{\bm \theta_i^+}-\left<H\right>_{ \bm \theta_i^-})/2,
\label{gradient}
\end{equation}
where $\bm \theta_i^{\pm} = \bm \theta \pm \frac{\pi}{2} \bm e_i$, $\bm e_i$ is the $i$-th unit vector in the parameter space. Higher order derivatives $\frac{\partial^{2} E(\bm \theta)}{\partial \theta^{2}_i}$, $\frac{\partial^{3} E(\bm \theta)}{\partial\theta^{3} _i}$, which are required in some optimizers, can be estimated in a similar way~\cite{mari2020estimating}.

\subsection{Experimental protocol}

\begin{figure}
 	\includegraphics[width=8cm]{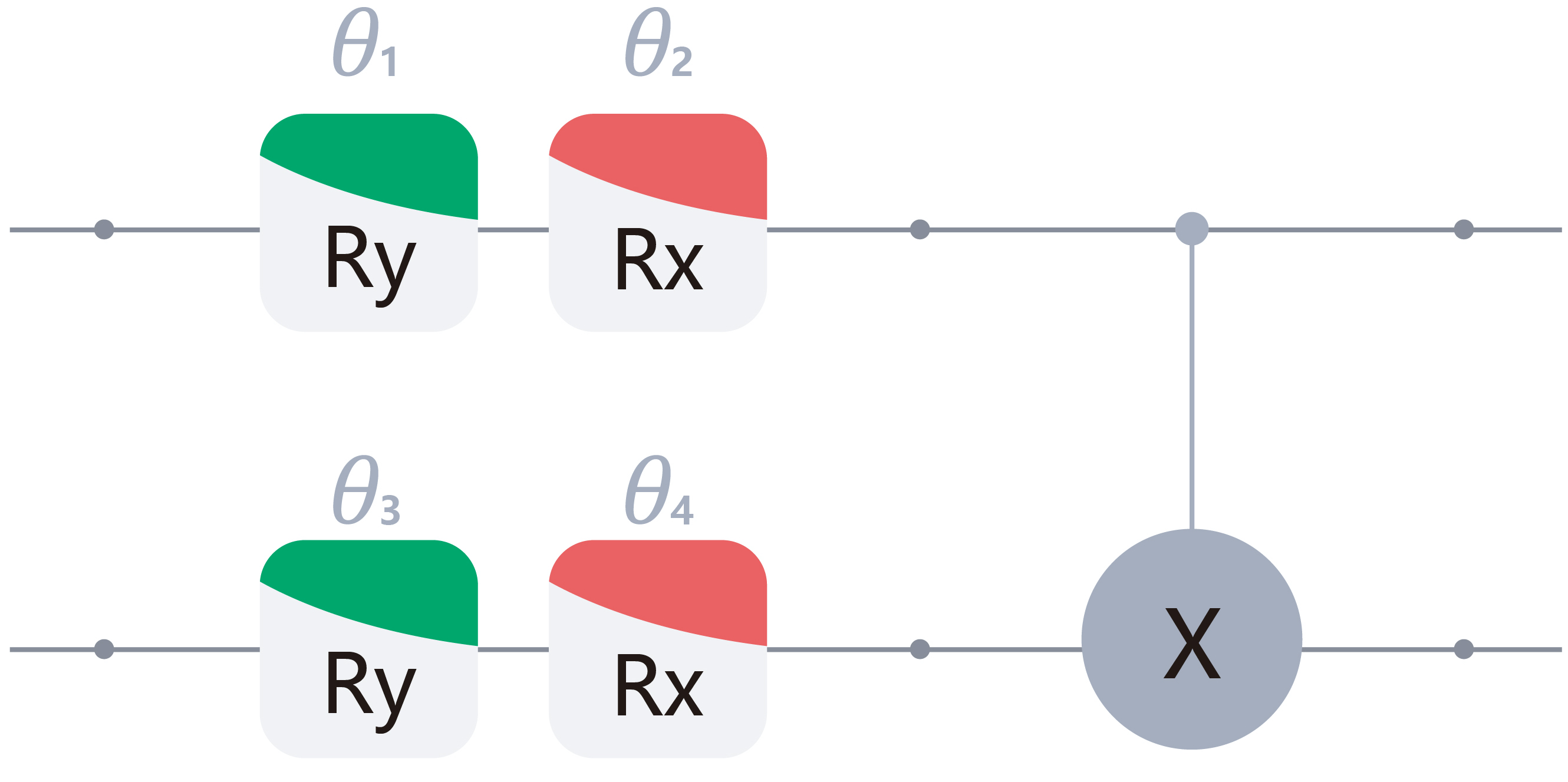}
 	\caption{The hardware efficient circuit for 2-qubit VQE. $\theta_1$ - $\theta_4$ are the parameters to be optimized.}
 	\label{ansatz}
 \end{figure}

In this work, we apply VQE  to find the ground state of 2-qubit Heisenberg model. The Hamiltonian is
\begin{equation}
H_{H} = X_1X_2 + Y_1 Y_2 + Z_1Z_2,
\end{equation} 
 where $X_j,Y_j,Z_j$ are the Pauli operator on the $j$-th qubit. The hardware efficient circuit is shown in Fig.~\ref{ansatz}. 
 
We implement experiments on SpinQ Gemini and IBM Q Yorktown with initial parameter $\bm \theta =\left[ 10.2^{\circ} , 8.35^{\circ}, 108^{\circ},91.5^{\circ}\right]$, learning rate $\alpha=0.25$, and carry out  numerical simulations. 

\begin{figure}
	\includegraphics[width=6cm]{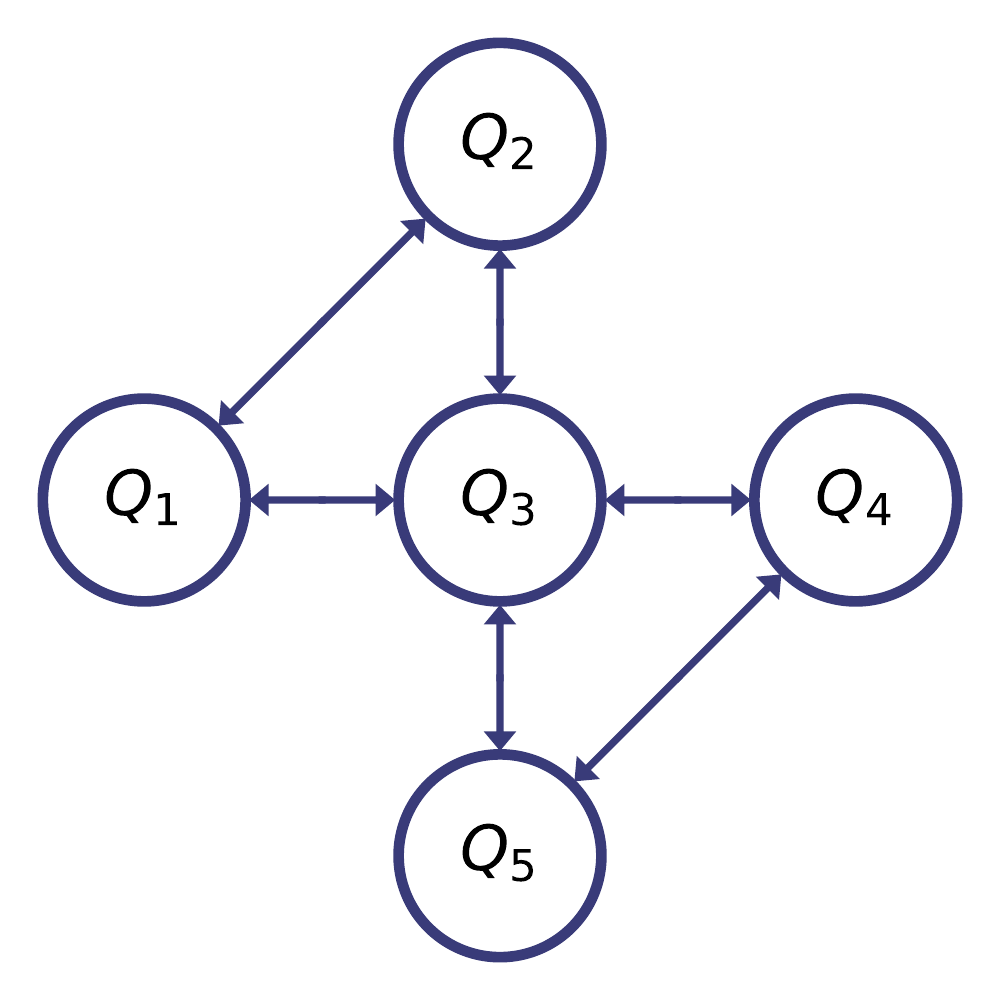}
	\caption{The hardware structure of IBM Q Yorktown.}
	\label{ibmq}
\end{figure}

IBM Q Yorktown is a superconducting quantum computer with 5 qubits~\cite{ibmq_5_yorktown}, the structure is shown in Fig.~\ref{ibmq}. We only use the first two qubits $Q_1$ and $Q_2$, the single-gate error rates are $1.173 \times 10^{-3}$ and $9.810 \times 10^{-4}$, the readout errors are $2.280 \times 10^{-2}$ and $3.660 \times 10^{-2}$, the CNOT error rate is $1.825 \times 10^{-2}$.

The experimental procedures are as follows:
\begin{itemize}
	\item Initialize the circuit parameters $\bm \theta$,
	\item Estimate the derivatives of $\bm \theta$  via parameter-shift rule, $\frac{\partial E(\bm \theta)}{\partial \theta_i}=(\left<H\right>_{\bm \theta_i^+}-\left<H\right>_{ \bm \theta_i^-})/2$. 
	\item Update the parameters with gradient descent; $\boldsymbol{\theta'}=\boldsymbol{\theta}-\alpha\cdot \nabla E(\boldsymbol{\theta})$;
	\item Estimate the expectation value $\langle \bm 0|U^{\dagger}(\bm \theta) H U(\bm \theta) | \bm 0\rangle$;
	\item Repeat steps 2-4 until convergence.
\end{itemize}

\subsection{Results and simulation}

\begin{figure*}[!htbp]
\includegraphics[scale=0.4]{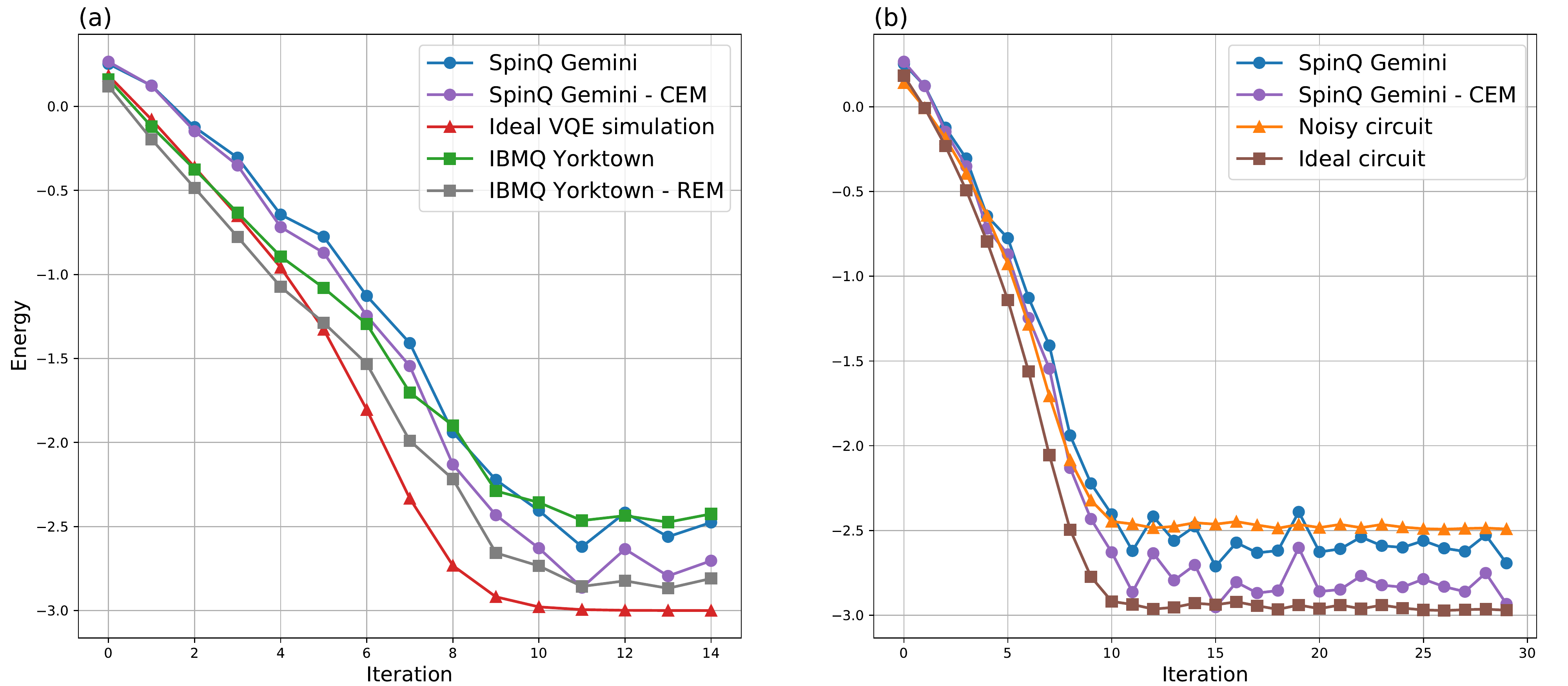}
\caption{Results of VQE.  (a) The VQE energy with respect to each iteration on different quantum computers and simulation. The blue dot line represents the experimental result on SpinQ Gemini. The purple dot line reperents the CNOT error mitigated(CEM) result on SpinQ Gemini. The red triangle line represents the numerical simulation without gate errors. The green square line represents the experimental result on IBM Q Yorktown. The gray square line represents the readout error mitigated(REM) result on IBM Q Yorktown.  (b) Numerical simulation for quantum circuit in VQE experiment. The blue dot line represents the experimental result on SpinQ Gemini. The purple dot line reperents the CNOT error mitigated result on SpinQ Gemini. The orange triangle line represents the energy calculate by the noisy circuit with the SpinQ Gemini experimental parameters. The brown square line represents the energy calculate by the ideal circuit with the SpinQ Gemini experimental parameters. } 
\label{vqe_result}
\end{figure*}

The results are shown in Fig.~\ref{vqe_result} (a).  show the original result of VQE experiment on SpinQ Gemini and IBMQ Yorktown, respectively. The ground state energy of $H_H$ is -3, which is shown by the ... line. SpinQ Gemini and IBM Q Yorktown perform similar, both converge to $E(\bm\theta) \approx -2.6$ after enough iterations, as shown by the ... line and the ... line, respectively.  According to our simulations and analysis, the error for Gemini mainly comes from the inhomogeneous of the magnetic field, while the error from IBMQ mainly comes from the readout error.

The noise in quantum computer can not be neglected. To study the noise effect and  stability of SpinQ Gemini, we construct a noise model to capture the quantum error of the SpinQ Genimi. In the realistic noise NMR quantum device, the basic noise channel are dephasing and amplitude damping noise. For initial state $\rho$ of the system and the quantum circuit unitary transformation $U$, the local noise model for single-qubit and two-qubit quantum gates  can be described by the Kraus representation

\begin{eqnarray}\label{noise_model}
\rho \rightarrow  \sum_{k} E_k U \rho U^{\dagger} E_k^{\dagger} = \sum_{k} E_k  \rho' E_k^{\dagger} ,
\end{eqnarray}
where $E_{k}$s are the Kraus operators and $\sum_{k} E_k E_k^{\dagger} = I$. The $E_k$s acting on the same single qubit and two qubits as $U$ acts on.  The amplitude damping noise can be characterized by the Kraus operators,
\begin{displaymath}
K_{1}=\left(
\begin{array}{cc}
1 & 0    \\
0  & \sqrt{1-p}   \\
\end{array}
\right), 
K_{2}=\left(
\begin{array}{cc}
0 & \sqrt{p}    \\
0  & 0  \\
\end{array}
\right),
\end{displaymath}
where $p \in [0, 1]$ is the probability of the noise.
For amplitude damping noise on single-qubit gate $U$, the Kraus operators $E_k$s in Eq.~\eqref{noise_model} run over the set $\{K_1, K_2\}$.
The Kraus operators $E_k$s run over the set $\{K_1, K_2\}\otimes\{K_1, K_2\}$ for two-qubit noisy gate.
The  dephasing noise is characterized by the Kraus operators,
\begin{eqnarray}
K_{1}=\sqrt{1-p}I_{2}, \quad  K_{2}=\sqrt{p}\sigma_{Z},
\end{eqnarray}
where $I_{2}$ is the two dimensional identity matrix and $\sigma_{Z}$ is Pauli operator. For dephasing  noise on single-qubit gate $U$, the Kraus operators $E_k$s in Eq.~\eqref{noise_model} run over the set $\{K_1, K_2\}$. The Kraus operators $E_k$s run over the set $\{K_1, K_2\}\otimes\{K_1, K_2\}$ for two-qubit noisy gate.

We model the noise consisting of single qubit thermal relaxation error and two qubit thermal relaxation error. The thermal relaxation error model applies the amplitude damping noise after dephasing noise in each one- or two-qubit gate. This  thermal relaxation error model is characterized through the parameters ($T_1, T_2^*,  t_{q}$) and the noise probability is formulated by
\begin{eqnarray}
p_{damping} &=& 1 - e^{-\frac{t_{q}}{T_1}}, \\
p_{dephasing} &=& \frac{1}{2}\left(1- e^{-2\gamma}\right), 
\end{eqnarray}
where $\gamma = \frac{t_q}{T_2^*} - \frac{t_q}{2T_1}$. When the thermal relaxation error model apply to single qubit gate, $t_{q} = {t_{1q}}$ and  $t_{q} = {t_{2q}}$ for two qubit gate.  The final noise model to approximate the noise of NMR quantum device is characterized by the parameters $\{T_1, T_2^*,  t_{1q}, t_{2q} \}$. We set $\{T_1 = 5.6s , T_2^*=0.025s,  t_{1q}=25\mu s, t_{2q}=800\mu s\}$ in the noise simulation for the NMR quantum computer. In NMR system, the dephasing effect is  caused by both the spin relaxation and the field inhomogeneity. $T_2$ is used to measure the spin relaxation rate, while $T_2^*$ is used to measure the field inhomogeneity.  The $T_2$ data is measured using the technique called spin echo, which can refocus the magnetisation and remove the effect of inhomogeneous field. In our VQE experiment, we did not use such technique, so we use $T_2^*$ instead of $T_2$.

With the noise model described above, we first record every parameters $\bm\theta$ in each iteration of  the SpinQ Gemini VQE experiment. Then we take these parameters $\bm \theta$ as the parameters of quantum circuit ansatz (Fig.~\ref{ansatz}) and calculate the energy of the Hamiltonian with respect to the ideal circuit and noisy circuit output in each iteration. As shown in Fig.~\ref{vqe_result} (b) , the noisy circuit result shows great consistancy to the experiment data. The paramaters $\bm \theta$ found by SpinQ Gemini is close to the parameters for ground state. These results indicate that our desktop quantum computer can run VQE algorithm well.

\subsection{Error mitigation}

Quantum error mitigation~\cite{li2017efficient, endo2018practical, barron2020measurement} is a technique to diminish the influence of errors from the statistical perspective. 

From the comparison and the simulation described above, we can see that the dephasing error caused by the inhomogeneous magnetic field is dominant. Our circuit consists of four single qubits rotations and one CNOT gate. The time for a CNOT gate is about 800 $\mu$s and for single qubit gates is $\sim 20$ $\mu$s. Therefore, the imperfections of the CNOT gate causes primary error. Consider the error model:
\begin{eqnarray}
\rho \rightarrow \rho_f =\sum_{k} E_k U \rho U^{\dagger} E_k^{\dagger} = \sum_{k} E_k  \rho' E_k^{\dagger},
\end{eqnarray}
where $E_k$'s are the Kraus operators, $\rho'$ is the ideal density matrix, and $\rho_f$ is the measured density matrix. Error mitigation is a procedure that for a given $\rho_f$ obtained from the experiment, find a density matrix $\rho_0$, which is as close to $\rho'$ as possible, so that the final experiment result could be improved. Here, we employ the superoperator formalism to obtain $\rho_0$. This formalism works as follows. First, let us rewrite the density matrix $\rho'$ from an $n\times n$ matrix into an $n^2 \times 1$ vector $\bm \rho'$:
\begin{equation}
\rho'=\sum_{ij} \rho'_{ij}|i\rangle\langle j| \rightarrow \bm \rho'=\sum_{ij} \rho'_{ij}|i\rangle |j\rangle.
\end{equation}
Then the final state $\bm \rho_f$, which is also an $n^2 \times 1$ vector is
\begin{equation}
\bm \rho_f = \hat{\hat{S}} \bm \rho',
\end{equation}
where $\hat{\hat{S}}$ is the superoperator. With known Kraus operators $E_k$, it can be obtained as 
\begin{equation}
\hat{\hat{S}}=\sum_k E_k\otimes E_k^{\dagger}.
\end{equation}
Therefore, with known $\bm \rho_f$ and $\hat{\hat{S}}$, we can get
\begin{equation}
\bm \rho'=\hat{\hat{S}}^{-1}\bm \rho_f.
\end{equation}
The original result and the mitigated result of Gemini is shown in \cref{vqe_result}(b). The blue line and the purple line show the original result and the CNOT error mitigated result. We can see that the error mitigated result of the ground state of $H_H$ could reach about $-2.98$, much closer to the ideal result. With both the simulation result and the error mitigated result, we can see that the error model we used is a good approximation.

For the IBMQ devices, the readout errors are dominant. Here we consider the simplest linear algebra measurement error mitigation scheme. On IBMQ Santiago we do projective measurement  and obtain one of the strings $\{0,1\}^{\otimes 2}$. Through tomography of measurement process, we get the probability of string $S_j$ becoming $S_k$, denoted by $P_{kj}$. Suppose we repeat the same measurement many times and have the string probability distribution $C_{\text {noisy }}$, then
\begin{equation}
C_{\text {mitigated}}=P^{-1} C_{\text {noisy }}
\end{equation}
provides the probability distribution with measurement error mitigated, although $P^{-1}$ is not a physical operation. Measurement error mitigation can efficiently improve the performance of VQE on IBMQ Santiago, as shown in Fig.~\ref{vqe_result}(a).

\section{Discussion}

For the next generations of SpinQ desktop quantum computer products, we will develop products running with more qubits (3$\sim$4). Currently, the design of a
$3$-qubit machine is underway and the product is expected to be released in the second quarter of 2021, with a comparable price as SpinQ Gemini (i.e. under $50$k USD). Along the way, compatible software modules with advanced pulse control functions will also be developed, providing more powerful abilities for quantum algorithm/control/error mitigation designs to meet the research needs of advanced users. Meanwhile, another direction is to make a simplified version of the current model, making it more portable and much lower cost (under $5$k USD). This simplified version is expected to be released in the fourth quarter of 2021, such that it can be more affordable for most K-12 schools around the world.

\section*{Acknowledgement} We thank Jun Li and Tao Xin for their contribution to the early stage of this project. 

\bibliography{vqe}

\end{document}